\documentclass[3p,twocolumn]{elsarticle}

\usepackage{amssymb}
\usepackage{natbib}

\journal{Journal of Crystal Growth}

\begin{document}

\begin{frontmatter}



\title{Polytetrahedral short-range order and crystallization stability in supercooled ${\rm Cu_{64.5}Zr_{35.5}}$ metallic liquid}


\author[imet,urfu,ihpp]{R.E.~Ryltsev\corref{cor1}}
\author[ihpp,mipt,urfu]{N.M. Chtchelkatchev}
\address[imet]{Institute of Metallurgy, Ural Division of Russian Academy of Sciences, Amudsena str. 101, 620016 Ekaterinburg, Russia}
\address[urfu]{Ural Federal University, Mira str. 19, 620002 Ekaterinburg, Russia}
\address[ihpp]{Institute for High Pressure Physics, Russian Academy of Sciences, 142190 Troitsk, Russia}
\address[mipt]{Moscow Institute of Physics and Technology, 141700, 9 Institutskiy per., Dolgoprudny, Moscow Region, Russia}
\cortext[cor1]{Corresponding author. E-mail: rrylcev@mail.ru}

\begin{abstract}
Development of reliable interatomic potentials is crucial for theoretical studies of relationship between chemical composition, structure and observable properties in glass-forming metallic alloys. Due to ambiguity of potential parametrization procedure, some crucial properties of the system, such as crystallization stability or symmetry of the ground state crystal phase, may not be correctly reproduced in computer simulations. Here we address this issue for ${\rm Cu_{64.5}Zr_{35.5}}$ alloy described by two modifications of embedded atom model potential as well as by \textit{ab initio} molecular dynamics. We observe that, at low supercooling, both models provide very similar liquid structure, which agrees with that obtained by \textit{ab initio} simulations. Hoverer, deeply supercooled liquids demonstrate essentially different local structure and so different crystallization stability. The system, which demonstrate more pronounced icosahedral sort-range order, is more stable to crystallization that is in agreement with Frank hypothesis.

\end{abstract}




\end{frontmatter}


\section{Introduction}

Understanding relationship between chemical composition, short-range order (SRO) and nucleation stability in supercooled metallic liquids is crucial to fabricate bulk metallic glasses. One of the generally accepted paradigm suggests that icosahedral SRO (ISRO) is responsible for stability of supercooled metallic liquids. Such SRO provides local minima of potential energy but it is incompatible with translational order that causes so-called geometric frustration~\cite{Tarjus2005JPCM,Berthier2011RevModPhys,Shintani2006NaturePhys}. Following the pioneering work by Frank \cite{Frank1952ProcRoySocLond}, years of research have established that this idea can explain dynamical arrest in supercooled liquids \cite{Royall2015PhysRep,Hallett2018NatureComm} as well as glass-forming ability of metallic alloys \cite{Cheng2009PRL,Soklaski2013PRB,Wu2013PRB,Wang2015JPhysChemA,Ryltsev2016JCP}.

Among other glass-forming metallic alloys the Cu-Zr-based ones are of special interest due to their high glass-forming ability~\cite{Xu2004ActMat,Wang2004AppPhysLett}. Moreover, simulated Cu-Zr alloys have become model systems demonstrating pronounced ISRO \cite{Cheng2009PRL,Soklaski2013PRB,Wu2013PRB,Wang2015JPhysChemA,Ryltsev2016JCP,Klumov2016JETPLett}.


Molecular dynamics (MD) is the main theoretical tool for investigating SRO in supercooled liquids and glasses because it provides direct information regarding atomic-level structure. The key point of any MD simulation is the choice of interaction potential determining all the system properties at microscopic level. The most relevant type of potentials for describing metallic alloys is Embedded Atom Model (EAM), which usually provides good description of structural properties of both liquid and glassy metallic alloys \cite{Mendelev2009PhilMag,Wilson2015PhilosMag,Klumov2018JCP}. However, development of EAM potentials is a rather complex task due to the fact that many independent parameters need to be determined. Such parameters are usually obtained via fitting procedure, during which some properties calculated by EAM match to those either experimentally measured or calculated by \textit{ab initio} methods. This procedure is often ambiguous one: some properties (especially those involved in fitting procedure) may be described rather well while others may not \cite{Vella2015JPhysChemB,Lad2017JCP}. In regard to glassforming systems that means that, despite of good description of structure, some crucial properties of the system, such as crystallization stability or symmetry of the ground state crystal phase, may not be correctly reproduced in simulations \cite{Ryltsev2018JCP,Klumov2018JCP}. For example, recently we showed that, despite of pronounced ISRO, ${\rm Cu_{64.5}Zr_{35.5}}$ alloy described by widely accepted EAM potential \cite{Mendelev2009PhilMag} nucleates in sufficiently lengthy simulations with the formation of $\rm Cu_2Zr$ intermetallic compound with structure of C15 Laves phase. Moreover the alloy is actually unstable to crystallization for large system sizes ($N > 20,000$).

Such problems can be in principle overcome within the framework of \textit{ab initio} molecular dynamics (AIMD). This method does not require any fitting and, in principle, can provide comprehensive information regarding structural properties of a system under consideration. However, strong limitations on time and spatial scales available in AIMD simulations raise concerns about applicability of the method for describing supercooled liquids and glasses.

  \begin{figure*}[h]
  \centering
  \includegraphics[width=0.9\textwidth]{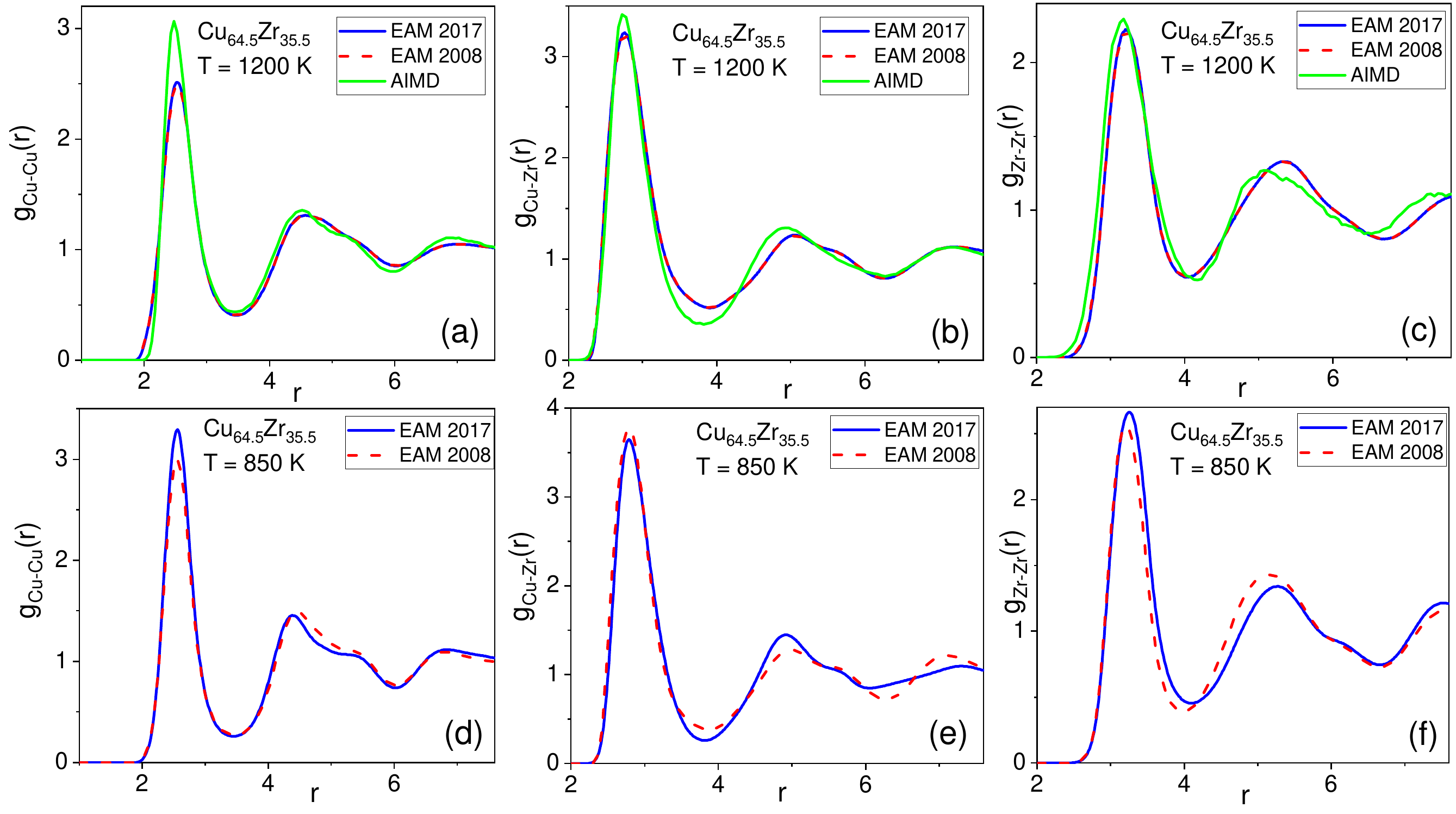}\\
  \caption{Comparison of radial distribution functions of ${\rm Cu_{64.5}Zr_{35.5}}$ alloy calculated by using two versions of EAM as well as by AIMD. (a)-(c) correspond to $T=1200$ (slightly supercooled liquid ) and (d)-(f) correspond to $T=850$ (deeply supercooled liquid near glass transition). }
  \label{fig:RDF}
\end{figure*}

Here we address these issues for  ${\rm Cu_{64.5}Zr_{35.5}}$ alloy described by two modifications of EAM potential as well as by AIMD. We show that, at low supercooling, both models provide very similar liquid structure, which is in good agrement with that obtained by AIMD. Hoverer, deeply supercooled liquids demonstrate essentially different SRO and so different crystallization stability. The system, which demonstrate more pronounced ISRO, is more stable to crystallization that is in agreement with Frank hypothesis.

\section{Methods}


For MD simulations, we use $\rm{LAMMPS}$ Molecular Dynamics Simulator \cite{Plimpton1995JCompPhys}. Periodic boundary conditions in Nose-Hoover NPT ensemble at $P=0$ were imposed. The number of particles was $N=50,000$.  The MD time step was 2 fs that provides good energy conservation at given thermodynamic conditions.

 Even though many properties of metals can be satisfactory described with using pair potentials \cite{Dubinin2011ThermochActa,Dubinin2014JNonCrystSol,Dzugutov1988PRA,Khusnutdinoff2018JETP}, many-body effects are usually important for describing crystallization (see, for example, Ref.~\cite{Baskes1999PRL}).  Thus, EAM potentials are applied in classical MD simulations of Cu-Zr alloys. We use two versions of EAM potential developed in the group of M. Mendelev from Ames Laboratory. The first one (hereinafter referred to as EAM-2008) is a widely accepted model \cite{Mendelev2009PhilMag,Zhang2015PRB}, which has been specially designed to describe liquid and glassy states of the Cu-Zr alloys. The second one (hereinafter referred to as EAM-2017) is modified version of EAM-2008 developed by the same group \cite{Mendelev_private}. The main difference between two models is due to pair interaction between Cu and Zr; the embedded function $F(\rho)$ for Zr and electron density functions $\rho(r)$ for Zr-Zr and Cu-Zr are also slightly different; other functions are the same for both models.

Initial configurations were prepared as hcp-lattices with random seeding of the species in the lattice sites. These configurations were melted, completely equilibrated at $T=1200$~K for 1 ns, then immediately cooled down to $T=850$~K and isothermally annealed at this temperature for $0.2$ $\mu$s. Note that annealing of $N=50,000$ particles for 0.2 $\mu$s ($\sim10^8$ MD steps) requires about a month of calculations in 128 supercomputer cores.

To study the structure of liquid, glassy and crystalline phases, we use radial distribution functions $g(r)$, Voronoi tessellation (VT) \cite{Finney1977Nature} and visual analysis of the snapshots. Detailed description of these methods is presented in Refs.~\cite{Ryltsev2015SoftMatt, Ryltsev2016JCP, Klumov2018JCP}.

The classical MD results for the high-temperature liquid state (at $T=1200$ K) were compared with those obtained by AIMD. To perform AIMD, we use open source quantum chemistry and solid state physics software package, CP2K. Projector augmented-wave (PAW) pseudopotentials and Perdew-Burke-Ernzerhof gradient approximation to the exchange-correlation functional were used. The simulations were performed at the $\Gamma$ point only, with a time step of 2 fs. Cubic supercell of 512 atoms with target composition  was built. Initial configuration was a random distribution of the atoms, which were subsequently equilibrated at 3000 K and 1200 K for 10 ps at each stage. Additional simulations for 10 ps were performed to collect data for calculating structural properties.

\section{Results}

Our first purpose is to compare two versions of EAM potential for Cu-Zr and to check how closely both models describe structure of the Cu-Zr liquids in comparison with AIMD. In Fig.~\ref{fig:RDF} we show partial radial distribution functions $g(r)$ of ${\rm Cu_{64.5}Zr_{35.5}}$ alloy calculated by using both versions of EAM as well as by AIMD. We see that both EAM models demonstrate practically the same structure at $T=1200$ K, which corresponds to slightly supercooled liquid (Fig.~\ref{fig:RDF}(a)-(c)). Comparison with AIMD reveals good agrement.  However, for deeply supercooled liquid at $T=850$ K, $g(r)$ for the EAM models under consideration are noticeably different (Fig.~\ref{fig:RDF}(d)-(f)).  Such difference is probably caused by difference in $U_{\rm Cu-Zr}$ which becomes important and low temperatures.

\begin{figure}
  \centering
  \includegraphics[width=0.8\columnwidth]{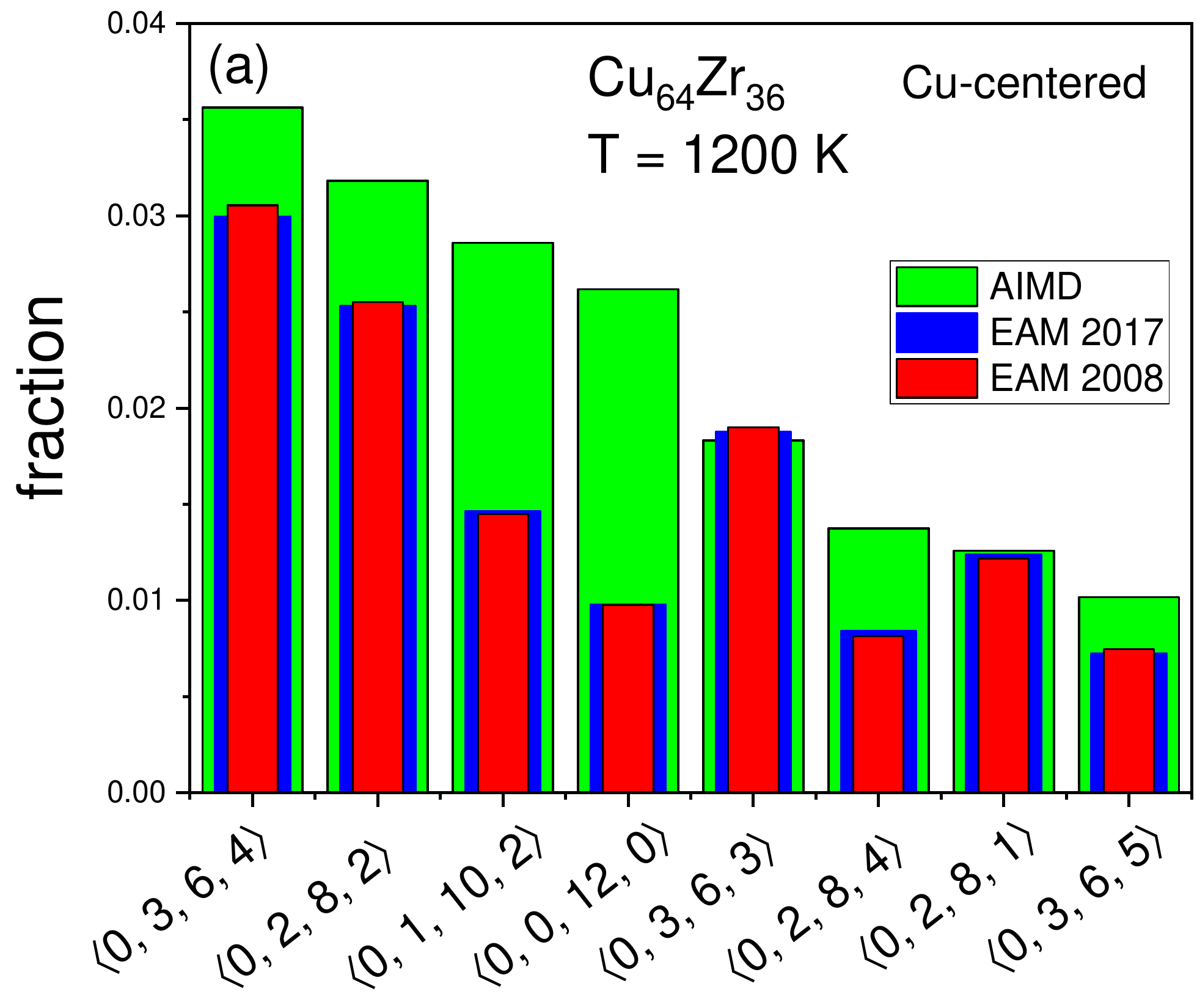}  \includegraphics[width=0.8\columnwidth]{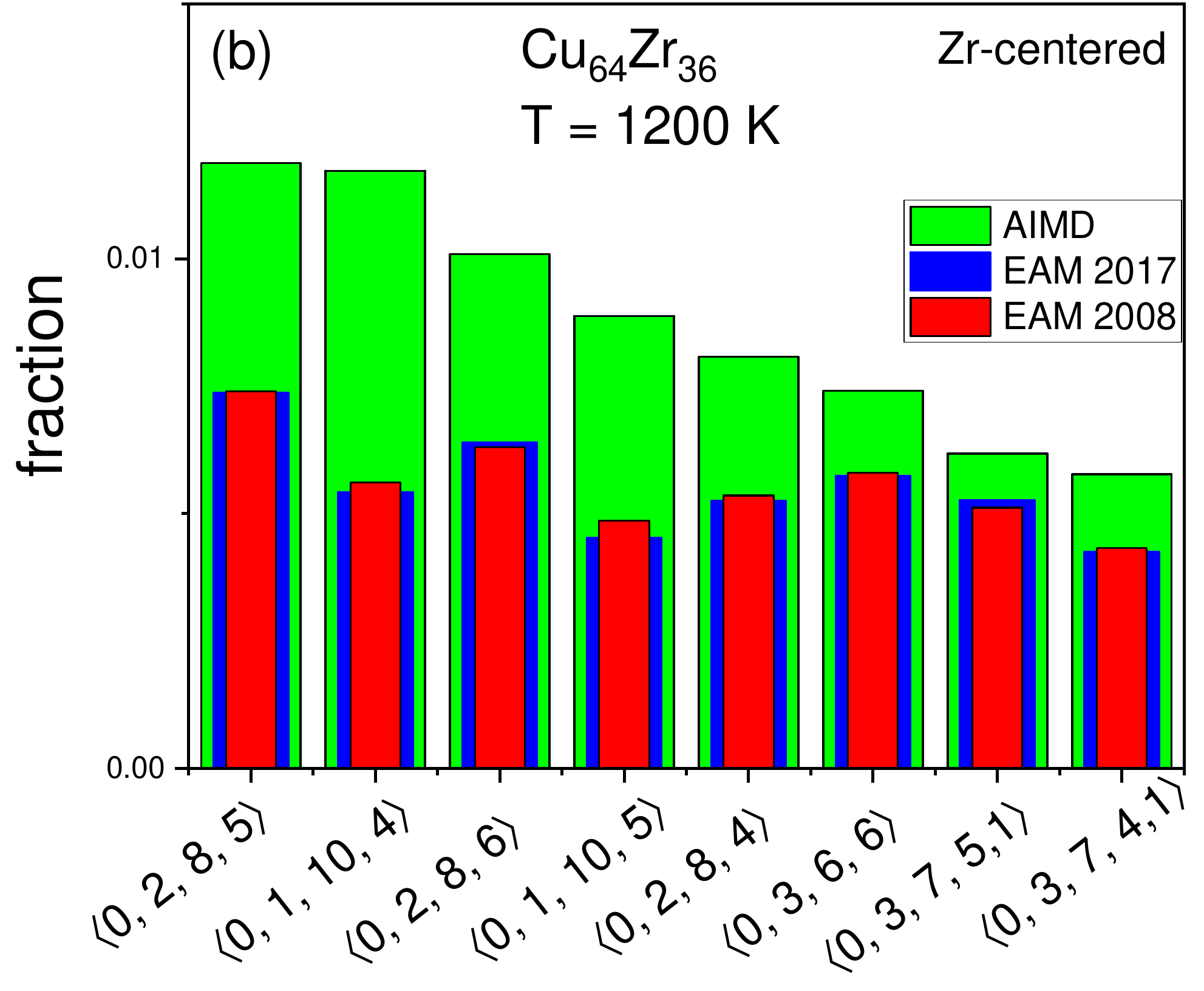}\\
  \caption{Histograms of the most popular Voronoi polyhedra for ${\rm Cu_{64.5}Zr_{35.5}}$ alloy at $T=1200$ K calculated by using two versions of EAM as well as by AIMD.}
  \label{fig:VT}
\end{figure}

In Fig.~\ref{fig:VT} we show histograms of the most popular Voronoi polyhedra for ${\rm Cu_{64.5}Zr_{35.5}}$ alloy at $T=1200$ K. We see again that both EAM potentials demonstrate practically the same results. Comparison with AIMD reveals qualitative agreement; the list of the most popular Voronoi polyhedra is the same for EAMs and AIMD but they fractions are essentially different. For example, fraction of icosahedra for AIMD is greater by a factor of two.

\begin{figure}
  \centering
  \includegraphics[width=\columnwidth]{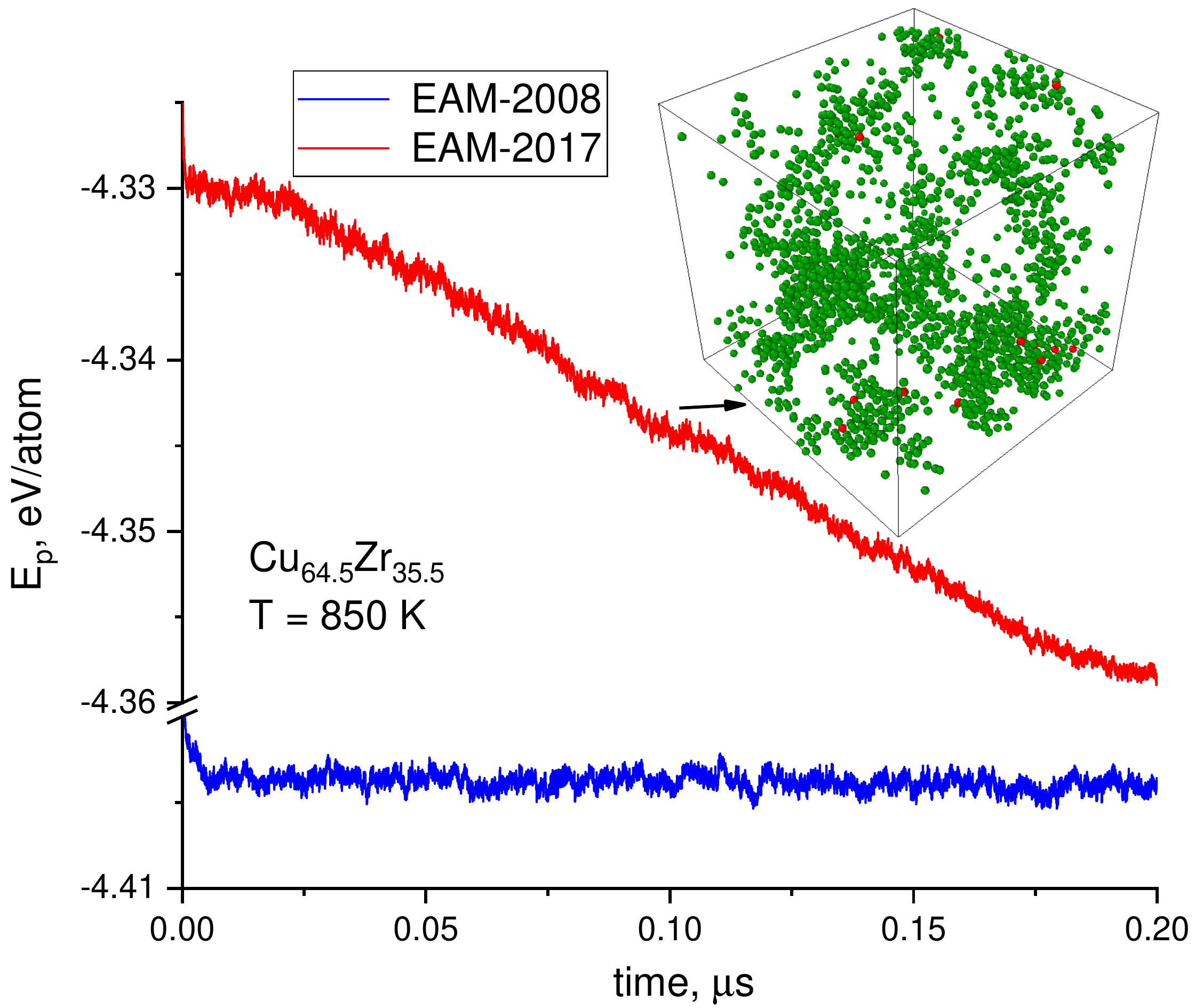}\\
  \caption{Time dependencies of potential energy $E_p$ for ${\rm Cu_{64.5}Zr_{35.5}}$ alloy described by different EAM potentials at $T=850$ K. Inset demonstrates spatial distributions of the centers of Z16 polyhedra at $t=0.1$ $\mu$s for EAM-2008.}
  \label{fig:Ept}
\end{figure}

Our main purpose it to study crystallization stability of simulated Cu-Zr alloys. Recently we reported that ${\rm Cu_{64.5}Zr_{35.5}}$ alloy described by EAM-2008 is actually unstable to crystallization for large system sizes ($N > 20,000$); it nucleates with the formation of Laves phases \cite{Ryltsev2018JCP}. Here we perform the same annealing for  ${\rm Cu_{64.5}Zr_{35.5}}$ with EAM-2017. In Fig.~\ref{fig:Ept}, we show time dependencies of the potential energy $E_{p}$ for both systems mentioned. We see that average $E_{p}$ is constant for EAM-2017 but decreases essentially for EAM-2008. That means the former is stable to crystallization on the simulation timescales.

\begin{figure}
  \centering
  \includegraphics[width=\columnwidth]{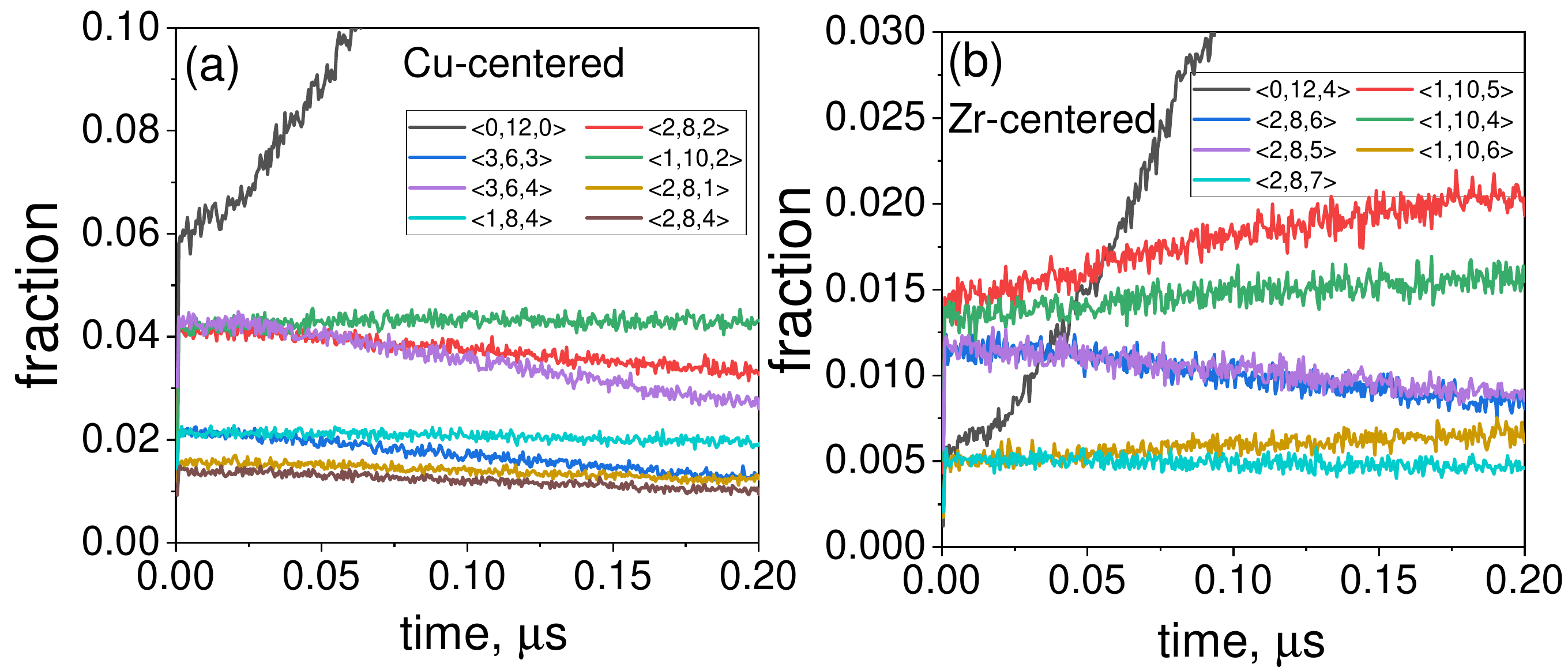} \includegraphics[width=\columnwidth]{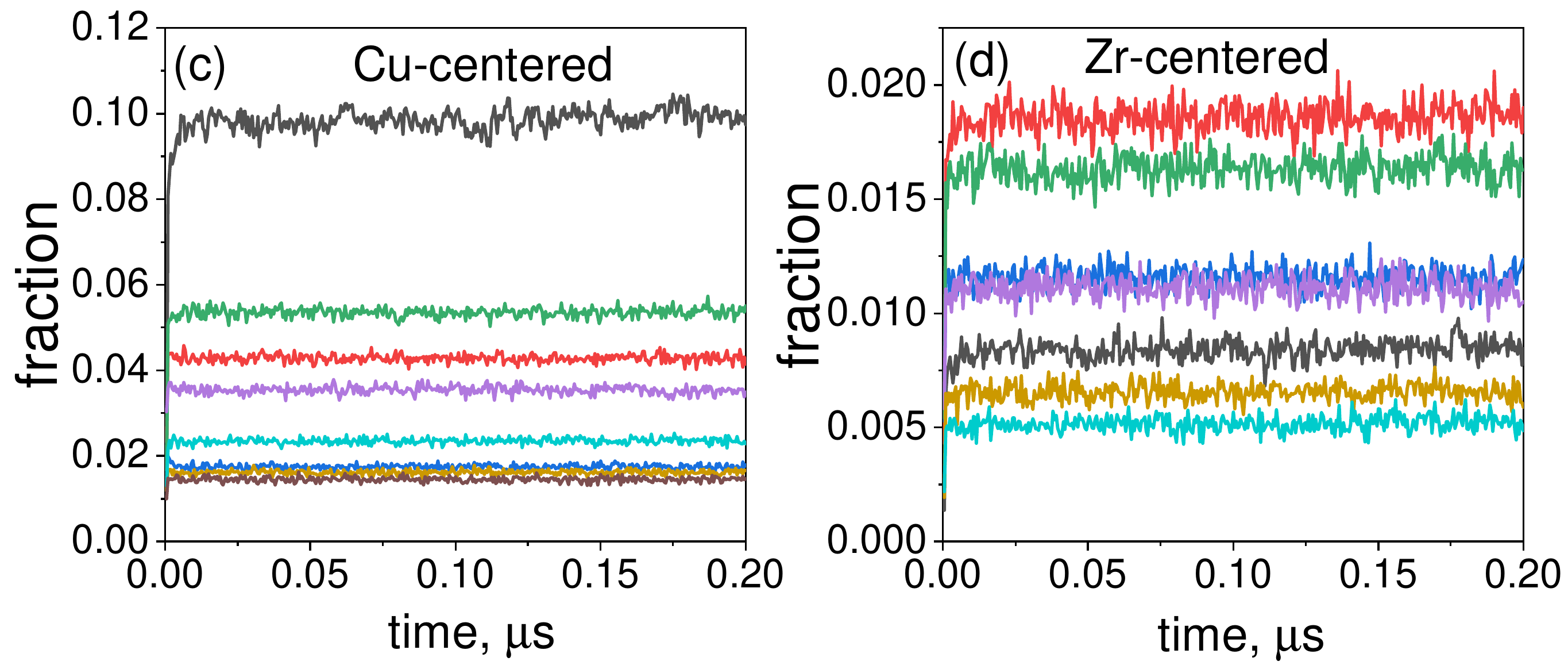} \\
  \caption{ Time dependencies of fractions of the most popular Voronoi polyhedra for ${\rm Cu_{64.5}Zr_{35.5}}$ alloy described by both EAM-2008 (a,b) and EAM-2017 (c,d) at $T=850$ K. Panels (a,c) and (b,d) correspond to Cu-centered and Zr-centered polyhedra, respectively.}
  \label{fig:VT_ann}
\end{figure}

To compare structural evolution of the systems under consideration, we perform Voronoi analysis. In Fig.~\ref{fig:VT_ann} we demonstrate time dependencies of the fractions of the most popular Voronoi polyhedra. All the polyhedra presented are ZCN Kasper polyhedra (including icosahedron Z12) and their distorted modifications (CN is coordination number) \cite{Cheng2011ProgMateSci}. Note that the increase of the fraction of both icosahedra and Z16 in ${\rm Cu_{64.5}Zr_{35.5}}$ with EAM-2017 is due to the growth of C15 Laves phase, which is build of these polyhedra \cite{Ryltsev2018JCP}. Comparison of the pictures reveals that  ${\rm Cu_{64.5}Zr_{35.5}}$ with EAM-2017 has more pronounced ISRO order than that with EAM-2008. Indeed the latter has only 6\% of icosahedral clusters in the supercooled liquid state (before the crystal growth starts) whereas the former has 10\% of icosahedra. We suggest that causes crystallization stability of ${\rm Cu_{64.5}Zr_{35.5}}$ EAM-2017 alloy that is in agreement with Frank hypothesis.


Analysis of Fig.~\ref{fig:VT_ann}(a,b) allows making important conclusions regarding growth of crystal nuclei in ${\rm Cu_{64.5}Zr_{35.5}}$ with EAM-2008. First, we see that fraction of Z16 polyhedra in supercooled liquid is order of magnitude less than fraction of icosahedra. That suggests the Z16 polyhedron is an useful indicator of the growth of Laves phases. In the insert for Fig.~\ref{fig:Ept} we show spatial distributions of the centers of Z16 polyhedra in ${\rm Cu_{64.5}Zr_{35.5}}$ alloy described by EAM-2008 at $T=850$ K and $t=0.1$ $\mu$s. This distribution clearly demonstrates the existence of several independent nuclei.  This observation is in agreement with recent studies \cite{Galimzyanov2018JETPLEtt,Galimzyanov2019ActaMater} where the authors observe that crystallization at low supercooling levels occurs through a mononuclear scenario but high concentration of crystal nuclei form at high supercooling levels.



\section{Conclusions}

Doing molecular dynamics simulations we compare structure and crystallization stability of two Embedded Atom Models for ${\rm Cu_{64.5}Zr_{35.5}}$ alloy developed by M. Mendelev and co-authors (EAM-2008 \cite{Mendelev2009PhilMag} and EAM-2017 \cite{Mendelev_private}). Exploring  their structural evolution at isothermal annealing near the glass transition temperature for 0.2 $\mu$s, we observe that crystallization stability of supercooled Cu-Zr liquids is closely related to ISRO that is in agreement with Frank hypothesis. Indeed, the alloy with EAM-2017 demonstrating more pronounced ISRO (about 10\% of ico-like clusters) is stable to crystallization at simulation timescales. Whereas, the one with EAM-2008 having about 6\% of ico-like clusters nucleates with the formation of C15 Laves phase. In general, SRO of Cu-Zr alloys is polytetrahedral that means it is mainly presented by Kasper polyhedra. Thus, besides icosahedra, other Kasper polyhedra, such as Z13 and Z11, can play important role in crystallization stability.

  We observe that systems demonstrating similar structure in both equilibrium and slightly supercooled liquid states can have noticeably different structure and crystallization stability being deeply supercooled. That conclusion may be important for developing interatomic potentials procedure. Indeed, our findings rise a question if an EAM potential developed with using fitting of high-temperature structure properties is suitable for simulating deeply supercooled liquids and glasses.

   We also observe that crystallization of the system at high supercooling level (i.e., at temperatures comparable to and below the glass transition temperature) occurs through  polynuclear scenario that is in agreement with recent studies \cite{Galimzyanov2018JETPLEtt,Galimzyanov2019ActaMater}.

\section{Acknowledgments}
The authors gratefully acknowledge M. Mendelev for helpful discussion and providing EAM potentials. This work was supported by Russian Science Foundation (grant RNF №18-12-00438). Molecular dynamic simulations have been carried out using  "Uran"\ supercomputer of IMM UB RAS and computing resources of the Federal collective usage center "Complex for Simulation and Data Processing for Mega-science Facilities"\ at NRC “Kurchatov Institute”.

\bibliographystyle{elsarticle-num-names}
\bibliography{our_bib}

\begin{thebibliography}{33}
\expandafter\ifx\csname natexlab\endcsname\relax\def\natexlab#1{#1}\fi
\providecommand{\url}[1]{\texttt{#1}}
\providecommand{\href}[2]{#2}
\providecommand{\path}[1]{#1}
\providecommand{\DOIprefix}{doi:}
\providecommand{\ArXivprefix}{arXiv:}
\providecommand{\URLprefix}{URL: }
\providecommand{\Pubmedprefix}{pmid:}
\providecommand{\doi}[1]{\href{http://dx.doi.org/#1}{\path{#1}}}
\providecommand{\Pubmed}[1]{\href{pmid:#1}{\path{#1}}}
\providecommand{\bibinfo}[2]{#2}
\ifx\xfnm\relax \def\xfnm[#1]{\unskip,\space#1}\fi
\bibitem[{Tarjus et~al.(2005)Tarjus, Kivelson, Nussinov, and
  Viot}]{Tarjus2005JPCM}
\bibinfo{author}{G.~Tarjus}, \bibinfo{author}{S.~A. Kivelson},
  \bibinfo{author}{Z.~Nussinov}, \bibinfo{author}{P.~Viot},
\newblock \bibinfo{title}{The frustration-based approach of supercooled liquids
  and the glass transition: a review and critical assessment},
\newblock \bibinfo{journal}{J. Phys.: Condensed Matt.} \bibinfo{volume}{17}
  (\bibinfo{year}{2005}) \bibinfo{pages}{R1143}.
\bibitem[{Berthier and Biroli(2011)}]{Berthier2011RevModPhys}
\bibinfo{author}{L.~Berthier}, \bibinfo{author}{G.~Biroli},
\newblock \bibinfo{title}{Theoretical perspective on the glass transition and
  amorphous materials},
\newblock \bibinfo{journal}{Rev. Mod. Phys.} \bibinfo{volume}{83}
  (\bibinfo{year}{2011}) \bibinfo{pages}{587--645}.
  \DOIprefix\doi{10.1103/RevModPhys.83.587}.
\bibitem[{Shintani and Tanaka(2006)}]{Shintani2006NaturePhys}
\bibinfo{author}{H.~Shintani}, \bibinfo{author}{H.~Tanaka},
\newblock \bibinfo{title}{Frustration on the way to crystallization in glass},
\newblock \bibinfo{journal}{Nature Phys.} \bibinfo{volume}{2}
  (\bibinfo{year}{2006}) \bibinfo{pages}{200}.
  \DOIprefix\doi{10.1038/nphys235}.
\bibitem[{Frank and Mott(1952)}]{Frank1952ProcRoySocLond}
\bibinfo{author}{F.~Frank}, \bibinfo{author}{N.~Mott},
\newblock \bibinfo{title}{Supercooling of liquids},
\newblock \bibinfo{journal}{Proceedings of the Royal Society of London. Series
  A. Mathematical and Physical Sciences} \bibinfo{volume}{215}
  (\bibinfo{year}{1952}) \bibinfo{pages}{43--46}.
  \DOIprefix\doi{10.1098/rspa.1952.0194}.
\bibitem[{Royall and Williams(2015)}]{Royall2015PhysRep}
\bibinfo{author}{C.~P. Royall}, \bibinfo{author}{S.~R. Williams},
\newblock \bibinfo{title}{The role of local structure in dynamical arrest},
\newblock \bibinfo{journal}{Phys. Reports} \bibinfo{volume}{560}
  (\bibinfo{year}{2015}) \bibinfo{pages}{1 -- 75}.
  \DOIprefix\doi{10.1016/j.physrep.2014.11.004}.
\bibitem[{Hallett et~al.(2018)Hallett, Turci, and
  Royall}]{Hallett2018NatureComm}
\bibinfo{author}{J.~E. Hallett}, \bibinfo{author}{F.~Turci},
  \bibinfo{author}{C.~P. Royall},
\newblock \bibinfo{title}{Local structure in deeply supercooled liquids
  exhibits growing lengthscales and dynamical correlations},
\newblock \bibinfo{journal}{Nature Communications} \bibinfo{volume}{9}
  (\bibinfo{year}{2018}) \bibinfo{pages}{3272}.
\bibitem[{Cheng et~al.(2009)Cheng, Ma, and Sheng}]{Cheng2009PRL}
\bibinfo{author}{Y.~Q. Cheng}, \bibinfo{author}{E.~Ma}, \bibinfo{author}{H.~W.
  Sheng},
\newblock \bibinfo{title}{Atomic level structure in multicomponent bulk
  metallic glass},
\newblock \bibinfo{journal}{Phys. Rev. Lett.} \bibinfo{volume}{102}
  (\bibinfo{year}{2009}) \bibinfo{pages}{245501}.
  \DOIprefix\doi{10.1103/PhysRevLett.102.245501}.
\bibitem[{Soklaski et~al.(2013)Soklaski, Nussinov, Markow, Kelton, and
  Yang}]{Soklaski2013PRB}
\bibinfo{author}{R.~Soklaski}, \bibinfo{author}{Z.~Nussinov},
  \bibinfo{author}{Z.~Markow}, \bibinfo{author}{K.~F. Kelton},
  \bibinfo{author}{L.~Yang},
\newblock \bibinfo{title}{Connectivity of icosahedral network and a
  dramatically growing static length scale in cu-zr binary metallic glasses},
\newblock \bibinfo{journal}{Phys. Rev. B} \bibinfo{volume}{87}
  (\bibinfo{year}{2013}) \bibinfo{pages}{184203}.
  \DOIprefix\doi{10.1103/PhysRevB.87.184203}.
\bibitem[{Wu et~al.(2013)Wu, Li, Wang, and Liu}]{Wu2013PRB}
\bibinfo{author}{Z.~W. Wu}, \bibinfo{author}{M.~Z. Li}, \bibinfo{author}{W.~H.
  Wang}, \bibinfo{author}{K.~X. Liu},
\newblock \bibinfo{title}{Correlation between structural relaxation and
  connectivity of icosahedral clusters in cuzr metallic glass-forming liquids},
\newblock \bibinfo{journal}{Phys. Rev. B} \bibinfo{volume}{88}
  (\bibinfo{year}{2013}) \bibinfo{pages}{054202}.
  \DOIprefix\doi{10.1103/PhysRevB.88.054202}.
\bibitem[{Wang et~al.(2015)Wang, Zhao, and Liu}]{Wang2015JPhysChemA}
\bibinfo{author}{D.~Wang}, \bibinfo{author}{S.-J. Zhao}, \bibinfo{author}{L.-M.
  Liu},
\newblock \bibinfo{title}{Theoretical study on the composition location of the
  best glass formers in cu-zr amorphous alloys},
\newblock \bibinfo{journal}{J. Phys. Chem. A} \bibinfo{volume}{119}
  (\bibinfo{year}{2015}) \bibinfo{pages}{806--814}.
  \DOIprefix\doi{10.1021/jp5120064}.
\bibitem[{Ryltsev et~al.(2016)Ryltsev, Klumov, Chtchelkatchev, and
  Shunyaev}]{Ryltsev2016JCP}
\bibinfo{author}{R.~E. Ryltsev}, \bibinfo{author}{B.~A. Klumov},
  \bibinfo{author}{N.~M. Chtchelkatchev}, \bibinfo{author}{K.~Y. Shunyaev},
\newblock \bibinfo{title}{Cooling rate dependence of simulated cu64.5zr35.5
  metallic glass structure},
\newblock \bibinfo{journal}{J. Chem. Phys.} \bibinfo{volume}{145}
  (\bibinfo{year}{2016}) \bibinfo{pages}{034506}.
  \DOIprefix\doi{10.1063/1.4958631}.
\bibitem[{Xu et~al.(2004)Xu, Lohwongwatana, Duan, Johnson, and
  Garland}]{Xu2004ActMat}
\bibinfo{author}{D.~Xu}, \bibinfo{author}{B.~Lohwongwatana},
  \bibinfo{author}{G.~Duan}, \bibinfo{author}{W.~L. Johnson},
  \bibinfo{author}{C.~Garland},
\newblock \bibinfo{title}{Bulk metallic glass formation in binary cu-rich alloy
  series cu100-xzrx (x=34, 36, 38.2, 40 at.\%) and mechanical properties of
  bulk cu64zr36 glass},
\newblock \bibinfo{journal}{Acta Mater.} \bibinfo{volume}{52}
  (\bibinfo{year}{2004}) \bibinfo{pages}{2621 -- 2624}.
  \DOIprefix\doi{10.1016/j.actamat.2004.02.009}.
\bibitem[{Wang et~al.(2004)Wang, Li, Sun, Sui, Lu, and
  Ma}]{Wang2004AppPhysLett}
\bibinfo{author}{D.~Wang}, \bibinfo{author}{Y.~Li}, \bibinfo{author}{B.~B.
  Sun}, \bibinfo{author}{M.~L. Sui}, \bibinfo{author}{K.~Lu},
  \bibinfo{author}{E.~Ma},
\newblock \bibinfo{title}{Bulk metallic glass formation in the binary cu-zr
  system},
\newblock \bibinfo{journal}{Appl. Phys. Lett.} \bibinfo{volume}{84}
  (\bibinfo{year}{2004}) \bibinfo{pages}{4029--4031}.
  \DOIprefix\doi{10.1063/1.1751219}.
\bibitem[{Klumov et~al.(2016)Klumov, Ryltsev, and
  Chtchelkatchev}]{Klumov2016JETPLett}
\bibinfo{author}{B.~A. Klumov}, \bibinfo{author}{R.~E. Ryltsev},
  \bibinfo{author}{N.~M. Chtchelkatchev},
\newblock \bibinfo{title}{Simulated cu--zr glassy alloys: the impact of
  composition on icosahedral order},
\newblock \bibinfo{journal}{JETP Lett.} \bibinfo{volume}{104}
  (\bibinfo{year}{2016}) \bibinfo{pages}{546--551}.
  \DOIprefix\doi{10.1134/S0021364016200017}.
\bibitem[{Mendelev et~al.(2009)Mendelev, Kramer, Ott, Sordelet, Yagodin, and
  Popel}]{Mendelev2009PhilMag}
\bibinfo{author}{M.~Mendelev}, \bibinfo{author}{M.~Kramer},
  \bibinfo{author}{R.~Ott}, \bibinfo{author}{D.~Sordelet},
  \bibinfo{author}{D.~Yagodin}, \bibinfo{author}{P.~Popel},
\newblock \bibinfo{title}{Development of suitable interatomic potentials for
  simulation of liquid and amorphous cu-zr alloys},
\newblock \bibinfo{journal}{Philos. Mag.} \bibinfo{volume}{89}
  (\bibinfo{year}{2009}) \bibinfo{pages}{967--987}.
  \DOIprefix\doi{10.1080/14786430902832773}.
\bibitem[{Wilson and Mendelev(2015)}]{Wilson2015PhilosMag}
\bibinfo{author}{S.~Wilson}, \bibinfo{author}{M.~Mendelev},
\newblock \bibinfo{title}{Anisotropy of the solid-liquid interface properties
  of the ni-zr b33 phase from molecular dynamics simulation},
\newblock \bibinfo{journal}{Philos. Mag.} \bibinfo{volume}{95}
  (\bibinfo{year}{2015}) \bibinfo{pages}{224--241}.
  \DOIprefix\doi{10.1080/14786435.2014.995742}.
\bibitem[{Klumov et~al.(2018)Klumov, Ryltsev, and
  Chtchelkatchev}]{Klumov2018JCP}
\bibinfo{author}{B.~A. Klumov}, \bibinfo{author}{R.~E. Ryltsev},
  \bibinfo{author}{N.~M. Chtchelkatchev},
\newblock \bibinfo{title}{Polytetrahedral structure and glass-forming ability
  of simulated ni–zr alloys},
\newblock \bibinfo{journal}{J. Chem. Phys.} \bibinfo{volume}{149}
  (\bibinfo{year}{2018}) \bibinfo{pages}{134501}.
  \DOIprefix\doi{10.1063/1.5041325}.
\bibitem[{Vella et~al.(2015)Vella, Stillinger, Panagiotopoulos, and
  Debenedetti}]{Vella2015JPhysChemB}
\bibinfo{author}{J.~R. Vella}, \bibinfo{author}{F.~H. Stillinger},
  \bibinfo{author}{A.~Z. Panagiotopoulos}, \bibinfo{author}{P.~G. Debenedetti},
\newblock \bibinfo{title}{A comparison of the predictive capabilities of the
  embedded-atom method and modified embedded-atom method potentials for
  lithium},
\newblock \bibinfo{journal}{The Journal of Physical Chemistry B}
  \bibinfo{volume}{119} (\bibinfo{year}{2015}) \bibinfo{pages}{8960--8968}.
  \DOIprefix\doi{10.1021/jp5077752}.
\bibitem[{Lad et~al.(2017)Lad, Jakse, and Pasturel}]{Lad2017JCP}
\bibinfo{author}{K.~N. Lad}, \bibinfo{author}{N.~Jakse},
  \bibinfo{author}{A.~Pasturel},
\newblock \bibinfo{title}{How closely do many-body potentials describe the
  structure and dynamics of cu–zr glass-forming alloy?},
\newblock \bibinfo{journal}{J. Chem. Phys.} \bibinfo{volume}{146}
  (\bibinfo{year}{2017}) \bibinfo{pages}{124502}.
  \DOIprefix\doi{10.1063/1.4979125}.
\bibitem[{Ryltsev et~al.(2018)Ryltsev, Klumov, Chtchelkatchev, and
  Shunyaev}]{Ryltsev2018JCP}
\bibinfo{author}{R.~E. Ryltsev}, \bibinfo{author}{B.~A. Klumov},
  \bibinfo{author}{N.~M. Chtchelkatchev}, \bibinfo{author}{K.~Y. Shunyaev},
\newblock \bibinfo{title}{Nucleation instability in supercooled cu–zr–al
  glass-forming liquids},
\newblock \bibinfo{journal}{J. Chem. Phys.} \bibinfo{volume}{149}
  (\bibinfo{year}{2018}) \bibinfo{pages}{164502}.
  \DOIprefix\doi{10.1063/1.5054631}.
\bibitem[{Plimpton(1995)}]{Plimpton1995JCompPhys}
\bibinfo{author}{S.~Plimpton},
\newblock \bibinfo{title}{Fast parallel algorithms for short-range molecular
  dynamics},
\newblock \bibinfo{journal}{J. Comput. Phys.} \bibinfo{volume}{117}
  (\bibinfo{year}{1995}) \bibinfo{pages}{1 -- 19}.
  \DOIprefix\doi{10.1006/jcph.1995.1039}.
\bibitem[{Dubinin et~al.(2011)Dubinin, Yuryev, and
  Vatolin}]{Dubinin2011ThermochActa}
\bibinfo{author}{N.~Dubinin}, \bibinfo{author}{A.~Yuryev},
  \bibinfo{author}{N.~Vatolin},
\newblock \bibinfo{title}{Straightforward calculation of the wca entropy and
  internal energy for liquid metals},
\newblock \bibinfo{journal}{Thermochimica Acta} \bibinfo{volume}{518}
  (\bibinfo{year}{2011}) \bibinfo{pages}{9 -- 12}.
  \DOIprefix\doi{https://doi.org/10.1016/j.tca.2011.01.041}.
\bibitem[{Dubinin et~al.(2014)Dubinin, Filippov, Yuryev, and
  Vatolin}]{Dubinin2014JNonCrystSol}
\bibinfo{author}{N.~Dubinin}, \bibinfo{author}{V.~Filippov},
  \bibinfo{author}{A.~Yuryev}, \bibinfo{author}{N.~Vatolin},
\newblock \bibinfo{title}{Excess entropy of mixing for binary square-well fluid
  in the mean spherical approximation: Application to liquid alkali-metal
  alloys},
\newblock \bibinfo{journal}{J. Non-Cryst. Solids} \bibinfo{volume}{401}
  (\bibinfo{year}{2014}) \bibinfo{pages}{101--104}.
  \DOIprefix\doi{10.1016/j.jnoncrysol.2014.01.046}.
\bibitem[{Dzugutov et~al.(1988)Dzugutov, Larsson, and
  Ebbsj\"o}]{Dzugutov1988PRA}
\bibinfo{author}{M.~Dzugutov}, \bibinfo{author}{K.-E. Larsson},
  \bibinfo{author}{I.~Ebbsj\"o},
\newblock \bibinfo{title}{Pair potential in liquid lead},
\newblock \bibinfo{journal}{Phys. Rev. A} \bibinfo{volume}{38}
  (\bibinfo{year}{1988}) \bibinfo{pages}{3609--3617}.
  \DOIprefix\doi{10.1103/PhysRevA.38.3609}.
\bibitem[{Khusnutdinoff et~al.(2018)Khusnutdinoff, Galimzyanov, and
  Mokshin}]{Khusnutdinoff2018JETP}
\bibinfo{author}{R.~M. Khusnutdinoff}, \bibinfo{author}{B.~N. Galimzyanov},
  \bibinfo{author}{A.~V. Mokshin},
\newblock \bibinfo{title}{Dynamics of liquid lithium atoms. pseudopotential and
  eam-type potentials},
\newblock \bibinfo{journal}{JETP} \bibinfo{volume}{126} (\bibinfo{year}{2018})
  \bibinfo{pages}{83--89}. \DOIprefix\doi{10.1134/S1063776118010041}.
\bibitem[{Baskes(1999)}]{Baskes1999PRL}
\bibinfo{author}{M.~I. Baskes},
\newblock \bibinfo{title}{Many-body effects in fcc metals: A lennard-jones
  embedded-atom potential},
\newblock \bibinfo{journal}{Phys. Rev. Lett.} \bibinfo{volume}{83}
  (\bibinfo{year}{1999}) \bibinfo{pages}{2592--2595}.
  \DOIprefix\doi{10.1103/PhysRevLett.83.2592}.
\bibitem[{Zhang et~al.(2015)Zhang, Wang, Mendelev, Zhang, Kramer, and
  Ho}]{Zhang2015PRB}
\bibinfo{author}{Y.~Zhang}, \bibinfo{author}{C.~Z. Wang},
  \bibinfo{author}{M.~I. Mendelev}, \bibinfo{author}{F.~Zhang},
  \bibinfo{author}{M.~J. Kramer}, \bibinfo{author}{K.~M. Ho},
\newblock \bibinfo{title}{Diffusion in a cu-zr metallic glass studied by
  microsecond-scale molecular dynamics simulations},
\newblock \bibinfo{journal}{Phys. Rev. B} \bibinfo{volume}{91}
  (\bibinfo{year}{2015}) \bibinfo{pages}{180201}.
  \DOIprefix\doi{10.1103/PhysRevB.91.180201}.
\bibitem[{Mendelev and Sun~et al.(????)}]{Mendelev_private}
\bibinfo{author}{M.~Mendelev}, \bibinfo{author}{Y.~Sun~et al.},
\newblock \bibinfo{journal}{private communications}  (????).
\bibitem[{Finney(1977)}]{Finney1977Nature}
\bibinfo{author}{J.~Finney},
\newblock \bibinfo{title}{Modelling the structures of amorphous metals and
  alloys},
\newblock \bibinfo{journal}{Nature} \bibinfo{volume}{266}
  (\bibinfo{year}{1977}) \bibinfo{pages}{309--314}.
  \DOIprefix\doi{10.1038/266309a0}.
\bibitem[{Ryltsev et~al.(2015)Ryltsev, Klumov, and
  Chtchelkatchev}]{Ryltsev2015SoftMatt}
\bibinfo{author}{R.~Ryltsev}, \bibinfo{author}{B.~Klumov},
  \bibinfo{author}{N.~Chtchelkatchev},
\newblock \bibinfo{title}{Self-assembly of the decagonal quasicrystalline order
  in simple three-dimensional systems},
\newblock \bibinfo{journal}{Soft Matter} \bibinfo{volume}{11}
  (\bibinfo{year}{2015}) \bibinfo{pages}{6991--6998}.
  \DOIprefix\doi{10.1039/C5SM01397F}.
\bibitem[{Cheng and Ma(2011)}]{Cheng2011ProgMateSci}
\bibinfo{author}{Y.~Cheng}, \bibinfo{author}{E.~Ma},
\newblock \bibinfo{title}{Atomic-level structure and structure-property
  relationship in metallic glasses},
\newblock \bibinfo{journal}{Prog. Mater. Sci.} \bibinfo{volume}{56}
  (\bibinfo{year}{2011}) \bibinfo{pages}{379 -- 473}.
  \DOIprefix\doi{10.1016/j.pmatsci.2010.12.002}.
\bibitem[{Galimzyanov et~al.(2018)Galimzyanov, Yarullin, and
  Mokshin}]{Galimzyanov2018JETPLEtt}
\bibinfo{author}{B.~N. Galimzyanov}, \bibinfo{author}{D.~T. Yarullin},
  \bibinfo{author}{A.~V. Mokshin},
\newblock \bibinfo{title}{Change in the crystallization features of supercooled
  liquid metal with an increase in the supercooling level},
\newblock \bibinfo{journal}{JETP Letters} \bibinfo{volume}{107}
  (\bibinfo{year}{2018}) \bibinfo{pages}{629--634}.
  \DOIprefix\doi{10.1134/S0021364018100089}.
\bibitem[{Galimzyanov et~al.(2019)Galimzyanov, Yarullin, and
  Mokshin}]{Galimzyanov2019ActaMater}
\bibinfo{author}{B.~N. Galimzyanov}, \bibinfo{author}{D.~T. Yarullin},
  \bibinfo{author}{A.~V. Mokshin},
\newblock \bibinfo{title}{Structure and morphology of crystalline nuclei
  arising in a crystallizing liquid metallic film},
\newblock \bibinfo{journal}{Acta Materialia} \bibinfo{volume}{169}
  (\bibinfo{year}{2019}) \bibinfo{pages}{184 -- 192}.
  \DOIprefix\doi{10.1016/j.actamat.2019.03.009}.

\end{thebibliography}

\end{document}